\documentclass{PoS}
\usepackage{amssymb,amsmath,graphicx}

\title{Probing TeV scale physics via ultra cold neutron decays and calculating non-standard baryon matrix elements}
\ShortTitle{Probing TeV scale physics via ultra cold neutron decays and calculations of non-standard baryon matrix elements}

\author{\speaker{Rajan Gupta,} \footnote{LA-UR-12-10238} \ \ Tanmoy Bhattacharya and Anosh Joseph 
\\
Theoretical Division, Los Alamos National Laboratory, Los Alamos, NM 87545, USA \\}
\author{Huey-Wen Lin and Saul D. Cohen\\
Department of Physics, University of Washington, Seattle, WA 98195\\}

\abstract{ We motivate undertaking precision analyses of neutron
  decays to look for signatures of new scalar and tensor interactions
  that can arise in extensions of the Standard Model at the TeV scale.
  The key ingrediant needed to connect experimental data with
  theoretical analysis are high-precision calculations of matrix
  elements of isovector bilinear operators between the decaying
  neutron and final state proton. We describe the status of our
  Lattice QCD program of using valence clover fermions
  on dynamical $N_f=2+1+1$ HISQ configurations generated by the MILC
  Collaboration. On the theoretical side we use the effective
  field theory method and provide both model independent
  and dependent analyses to obtain bounds on possible
  scalar and tensor interactions, both from low energy experiments and
  LHC data. }

\FullConference{The XXIX International Symposium on Lattice Field Theory\\
July 10-16, 2011\\
Lake Tahoe, California\\
}

\begin{document}


\section{Introduction}

New physics at a given high energy scale, such as that generated by
the exchange of any heavy boson, gives rise to measurable effects at
the hadronic scale.  The most well known example is weak interactions
(WI) mediated by the exchange of $W$ and $Z$ vector bosons that are
responsible, for example, for neutron $\beta$-decay.  In the 1930s,
Fermi parameterized the charged current decay in terms of an effective
four-fermion interaction.  This effective theory was also able to
explain muon decay in terms of a single coupling constant $G_F$.
Elucidation of the Lorentz structure of the interaction, $V-A$, came
in 1957 with the work of Marshak and Sudershan, and Feynman and
Gell-Mann. Until then, physicists had considered scalar and tensor
interactions as viable candidates.  Analysis of low energy decays
revealed a number of properties of WI, such as quark-lepton
universality, and the weak scale, $\sim 100$ GeV, related to the mass
of the $W$ and $Z$ vector bosons that are now part of the electroweak
theory. In formulating a consistent electroweak theory, it was also
clear that associated with the weak scale there has to be a second
scale, $O({\rm TeV})$, characterizing the breaking of electro-weak
symmetery since at the hadronic scale WI are $10^{-5}$ times smaller
than the electric force.

To break the electro-weak symmetery, explain the origin of masses of
quarks, leptons and weak bosons (Higgs phenomena) and to stabilize the
Standard Model against radiative corrections motivates the need for
new physics at the TeV scale.  The LHC has, in fact,  been designed to
probe new physics at this scale. To explore the $1-10$ TeV range,
and re-examine the possibility of scalar and tensor interactions, we
follow the complementary approach to direct detection at the LHC by
looking for deviations from the SM in precision experiments at the
hadronic scale.  This talk gives an overview of the combined
experimental and theoretical precision analyses of ultracold neutron
(UCN) decays required to probe novel interactions at the TeV Scale. In
particular, we would like to address the following questions:
\begin{itemize}
\item
Are there novel scalar or tensor interactions at the TeV scale?
\item
What are the respective coupling constants and what bounds can be put 
on them?
\item
What are their manifestations at the hadronic scale and what decays/processes
are the best candidates for investigating them or for setting upper
bounds on their strengths?
\item
What beyond the SM candidate theories can accomodate these
new interactions and satisfy all other known constraints?
\end{itemize}

Decays of UCN are interesting for probing new physics because there
are terms in the neutron decay distribution that are particularly
sensitive to scalar and tensor interactions at the $10^{-3}$ level.
In these terms in the decay distribution, the SM contributions are
suppressed due to the helicity flip factor $m_e/E_e$, so they are of
the same order as the signal. The details of the theoretical
framework, analysis and preliminary results are given in
Ref.~\cite{Bhattacharya:2011}, and we will follow the notation
established there. In this talk we summarize the key points of the
formalism and give some preliminary results. A discussion of the
Lattice QCD part of the project will be presented by
Lin~\cite{HWLin:lat2011} and of the exisiting bounds from other low
energy experiments and from the LHC by
Bhattacharya~\cite{Bhattacharya:lat2011}.

\section{Neutron decay}

The differential neutron decay distribution function $D(E_e,
\mathbf{p} _e, \mathbf{p} _\nu, \boldsymbol{\sigma} _n)$, keeping only
terms relevant to the discussion here, can be written
as~\cite{Abele:2008zz,Bhattacharya:2011}
\begin{equation}
D (E_e, \mathbf{p} _e, \mathbf{p} _\nu, \boldsymbol{\sigma} _n)  = 
1 +  c_0  + c_1 \, \frac{E_e}{M_N}   + \frac{m_e}{E_e} \bar{b}  + 
 \bar{B} (E_e)   \frac{\boldsymbol{\sigma}_n \cdot \mathbf{p }_\nu}{E_\nu} + \cdots ~, 
\label{eq:diffn}
\end{equation}
where $ \mathbf{p} _e$ and $\mathbf{p} _\nu$ denote the electron and
neutrino three-momenta, and $\boldsymbol{\sigma} _n$ denotes the
neutron polarization. Here $\bar{b} \equiv b^{SM} + B^{BSM}$ is an effective Fierz
interference term and $\bar{B} (E_e)$ is an effective energy-dependent
correlation coefficient given by 
\begin{equation}
\bar{B} (E_e) =  B_{\rm LO} (\tilde{\lambda}) + c_0^{(B)} +  
c_1^{(B)} \frac{E_e}{M_N}  +  \frac{m_e}{E_e}  \left( b_\nu^{\rm SM} +   b_\nu^{\rm BSM}  \right)
\end{equation}
The terms most sensitive to novel scalar and tensor interactions are
proportional to the helicity-flip factor $\frac{m_e}{E_e}$, and
characterized by the parameters $\bar{b}$ and $b_\nu^{\rm SM} +
b_\nu^{\rm BSM}$.  The beyond the Standard Model (BSM) contributions
to these parameters arise from novel scalar and tensor interactions
(with coupling constants $\epsilon_S$ and $\epsilon_T$) generated at
the TeV scale.  To extract $\bar{b}$ requires precison measurements of
the electron spectrum, while $(b_\nu^{\rm SM}+b_\nu^{\rm BSM})$
requires measuring the asymmetry between the neutron polarization and
the anti-neutrino momentum.

Current bounds on $b^{BSM}$ and $b_\nu^{\rm BSM}$ leave open the
possibility that these could be as large as $10^{-3}$, which implies
that the scale for these novel interactions is TeV. Standard Model
contributions, $b^{\rm SM}$ and $b_\nu^{\rm SM}$, are $O(10^{-3})$ and
known to $O(10^{-5})$. So the sought after signal is the same size as
the well-understood SM background. Two experiments at the Los Alamos
ultracold neutron source, called UCNb~\cite{UCNb} and
UCNB~\cite{WilburnUCNB}, are being carried out by our experimental
colleagues to measure these parameters at the $10^{-3}$ level.  In a
UCN decay the recoil effects, $(p_n-p_p)/M_N \sim 10^{-3}$, are small
and of the same size as the signal while there are no nuclear
effects. Even though the signal in UCN decays is not swamped
by the background, nevertheless, the experiments are very challenging.

\section{UCNb and UNCB experiments}

\begin{figure}
\includegraphics[height=.23\textheight]{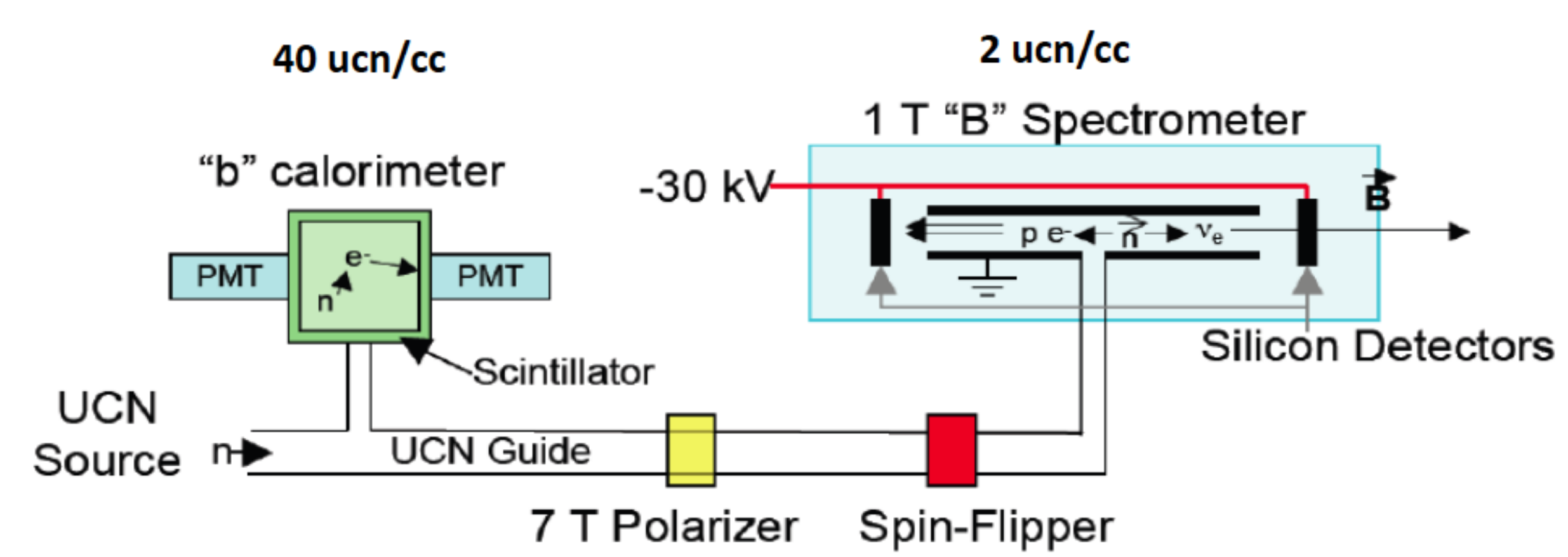}
\caption{
\label{Fig:exp}
The experimental set up for UCNb (left panel) and UCNB (right panel). 
The UCNb experiment is a calorimeter that measures the 
spectrum of the electron.  In the UCNB experiment the UCN enter the
spectrometer after being polarized to over $99.5\%$. The asymmetry
between the momentum direction of the neutrino and the spin of the
decaying neutron is measured indirectly by detecting
coincidences between the proton and the electron. The analysis is 
more complex and what one extracts in the end is 
$\bar{b} - (b_\nu^{\rm SM} + b_\nu^{\rm BSM})$. }
\end{figure}

The two separate experiments, UCNb and UNCB, will use the same UCN
source.  The setup is shown in Fig.~\ref{Fig:exp}. The proposed plan is to
make $10^{-3}$ measurements of $\bar{b}$ and $\bar{b}-(b_\nu^{\rm SM}
+ b_\nu^{\rm BSM})$ by the end of 2013 and based on the results and
detector performance and sensitivity achieved propose a $10^{-4}$
level experiment to DOE.

\begin{figure}
\includegraphics[width=.99\textwidth]{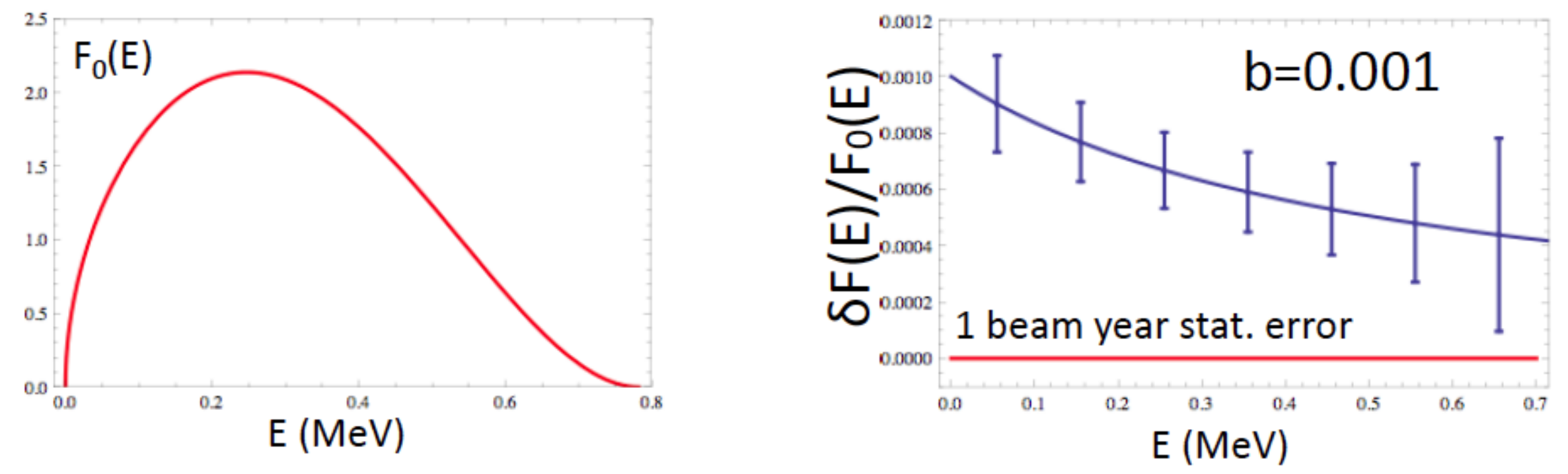}
\caption{
\label{Fig:b}
The left panel shows a plot of the electrum spectrum $F_0(E)$.  The
right panel shows the expected deviations assuming $\bar{b}=0.001$, and the 
statistical errors with one year of running.  }
\end{figure}

The goal of the UCNb experiment is a high precision measurement of the
electron decay spectrum. In Fig.~\ref{Fig:b} we show the full SM
expected spectrum $F_0(E_e)$ of the electron in the limit of zero
nuclear recoil (left panel) and the deviation from it assuming
$b^{BSM}=10^{-3}$ (right panel).  To achieve the design sensitivity
requires building a very sensitive and high resolution calorimeter.
The UCNB experiment measures the asymmetry between the momentum
direction of the neutrino and the spin of the decaying neutron and
yields $\bar{b} - (b_\nu^{\rm SM} + b_\nu^{\rm BSM})$. The neutrino momentum
is determined indirectly by measuring the electron and proton momenta
in coincidence. (In practice, the precision possible on these momenta
is not sufficient to fully reconstruct the neutrino momentum so a
reduced-detail observable is used from which $\bar{b} - (b_\nu^{\rm
 SM} + b_\nu^{\rm BSM})$ will be extracted~\cite{Bhattacharya:2011}.)

\section{Existing constraints}

\begin{figure}[!b]
\centering
\includegraphics[width=0.45\textwidth]{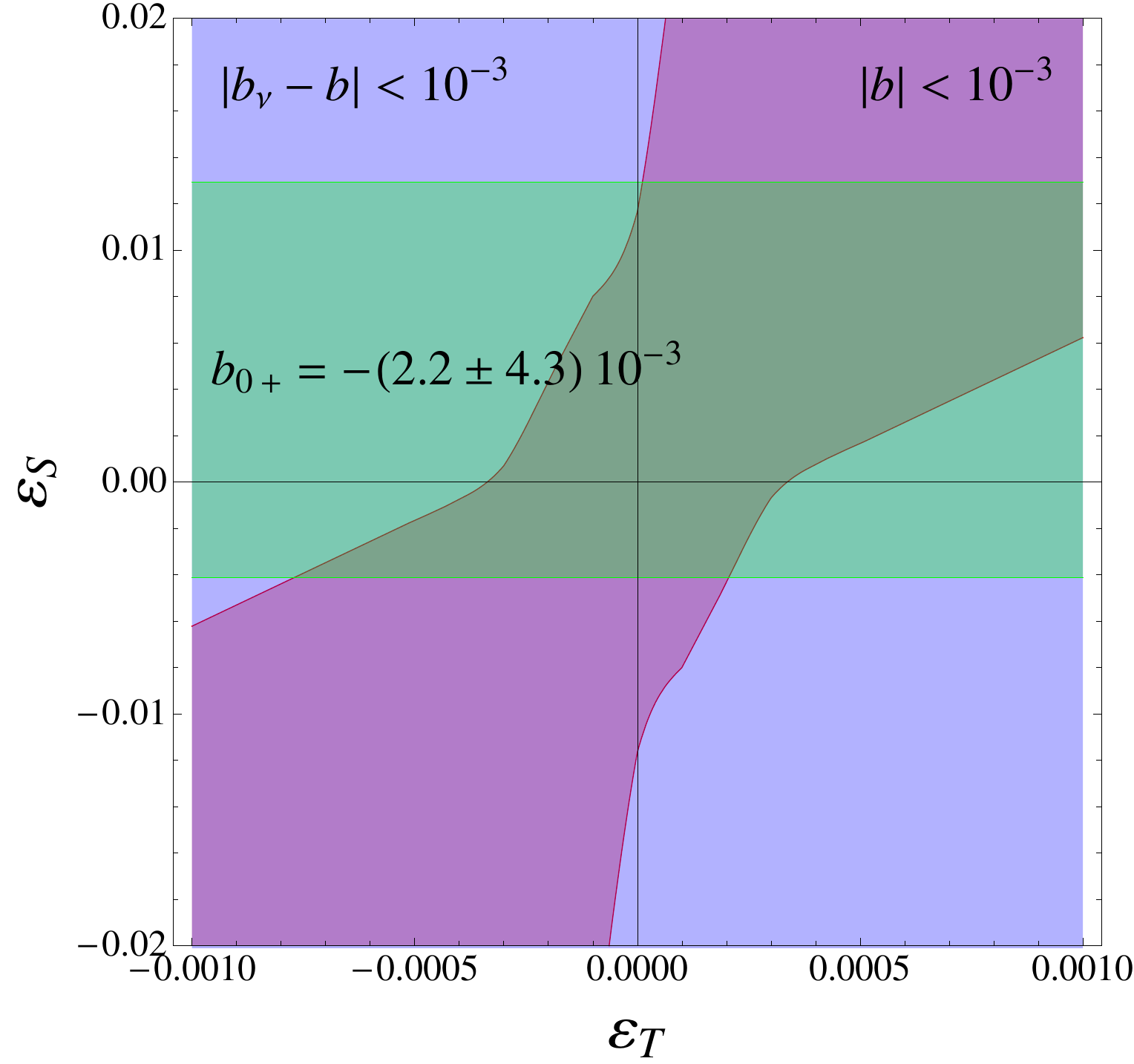}
\hspace{0.2in}
\includegraphics[width=0.45\textwidth]{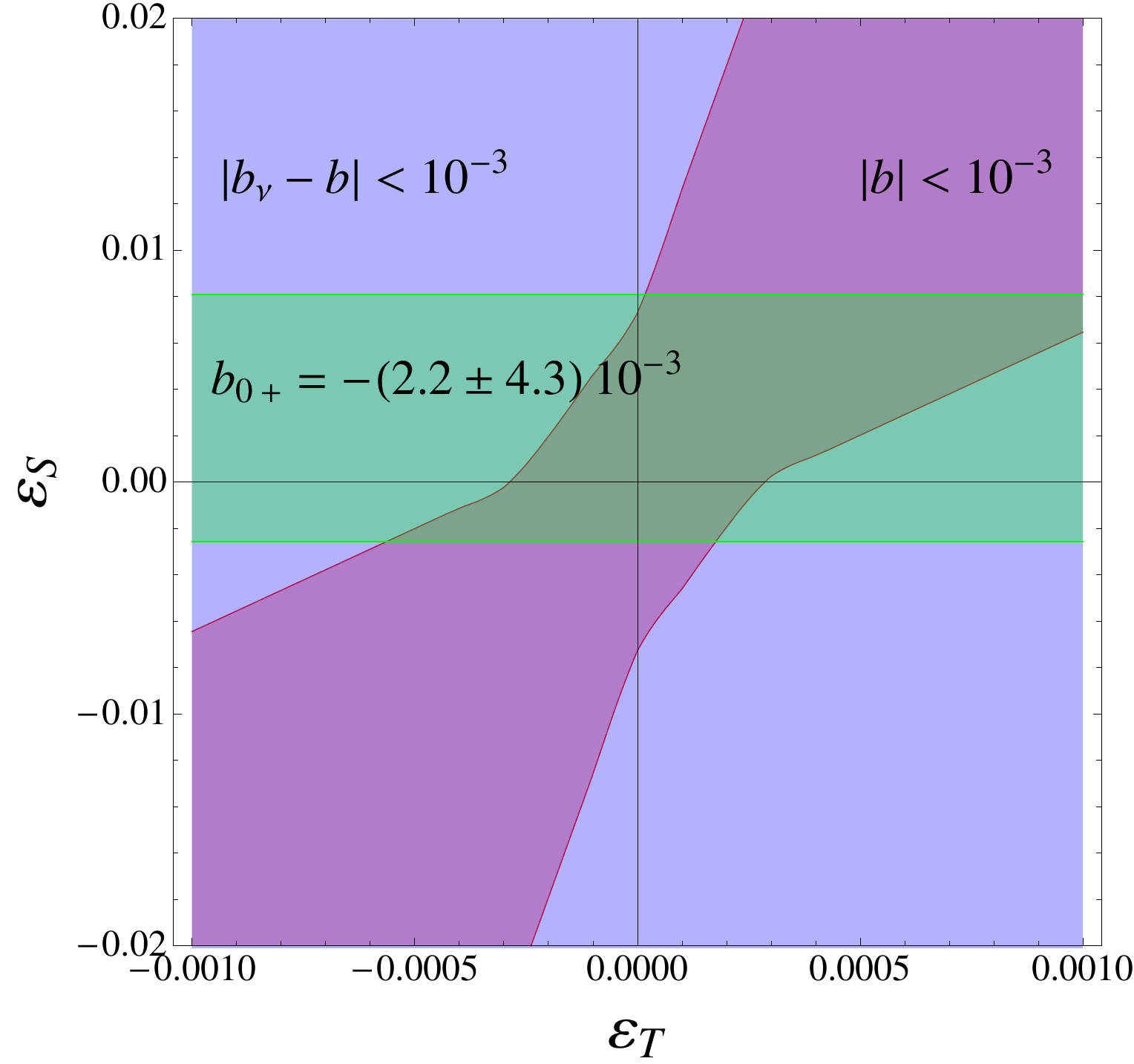}
\caption{Left panel: $90\%$ Confidence Level (C.L.) allowed regions in
  the $\epsilon_S$-$\epsilon_T$ plane implied by (i) the existing
  bound on $b_{0^+}$ characterizing the $0^+ \to 0^+$ nuclear beta
  decays (green horizontal band); (ii) projected measurements of
  $\bar{b}$ and $(b_\nu^{\rm SM} + b_\nu^{\rm BSM})-\bar{b}$ in
  neutron decay (red bow-tie shapes and blue regions) at the $10^{-3}$
  level; (iii) hadronic matrix elements taken in the ranges $0.25 <
  g_S < 1.0$, $0.6 < g_T < 2.3$~\cite{Herczeg2001vk}.  Right panel:
  same as left panel but with scalar and tensor charges taken from
  Lattice QCD: $g_{S}= 0.8(4)$ and $g_{T}=
  1.05(35)$~\cite{Bhattacharya:2011}.  Note that the constraint from
  $b_{0^+}$ also improves with higher precision in $g_S$. The
    effective couplings $\epsilon_{S,T}$ are defined in the
    $\overline{\rm MS}$ scheme at 2~GeV.
\label{fig:STconstraints}
}
\end{figure}

In Fig.~\ref{fig:STconstraints}, we illustrate how precision measurements of
$b$ and $b_\nu$ at the $10^{-3}$ level can reveal the existence of new
physics with mass scale $\Lambda_i$ in the multi-TeV range, which will
be explored directly at the LHC. Furthermore, the two panels highlight
the difference in bounds between using previous phenomenological
estimates for $g_S$ and $g_T$~\cite{Herczeg2001vk} and current LQCD 
estimates~\cite{Bhattacharya:2011,HWLin:lat2011}. In these
figures, $b_{0^+}$ represents the existing constraint from $0^+ \to 0^+$
nuclear beta decays.  While the LHC will be able to produce $W^{\rm
  BSM}$ with masses in the 0.2--3~TeV range and may provide precise
measurements of their masses, the proposed neutron decay experiments will
provide the most precise measurement of their spin and couplings,
needed to reconstruct the TeV-scale theory that will replace the
SM. Moreover, even if $W^{\rm BSM}$ is out of the kinematic reach of the
LHC (couplings are $O(10^{-4})$ or smaller) or swamped by background,
the interactions that it generates may lead to observable effects in
the proposed neutron-decay experiments.

To connect the new theory (the theory replacing the SM at the TeV
scale) to low-energy experiments requires the calculation of the
matrix elements of the new interactions between the decaying neutron
and the final state proton. Even through the new theory is unknown,
its consequences at the hadronic scale can be evaluated in terms of an
effective theory. In this model-independent effective theory, the
leading operators encapsulating novel scalar and tensor interactions
are iso-vector quark bilinear operators $\overline{u} d$ and
$\overline{u} \sigma^{\mu\nu}d$.  Knowing their matrix elements at $q
\equiv p_n - p_p = 0$, $i.e.$ the scalar and tensor charges $g_S$ and
$g_T$, will allow us to put bounds on the couplings
$\epsilon_{S,T}$. The best method for calculating these matrix
elements with control over all systematic errors is Lattice QCD. An
overview of our ongoing lattice calculations is given next.

\section{Lattice QCD Calculations}

\begin{figure}
\begin{center}
\includegraphics[height=.23\textheight]{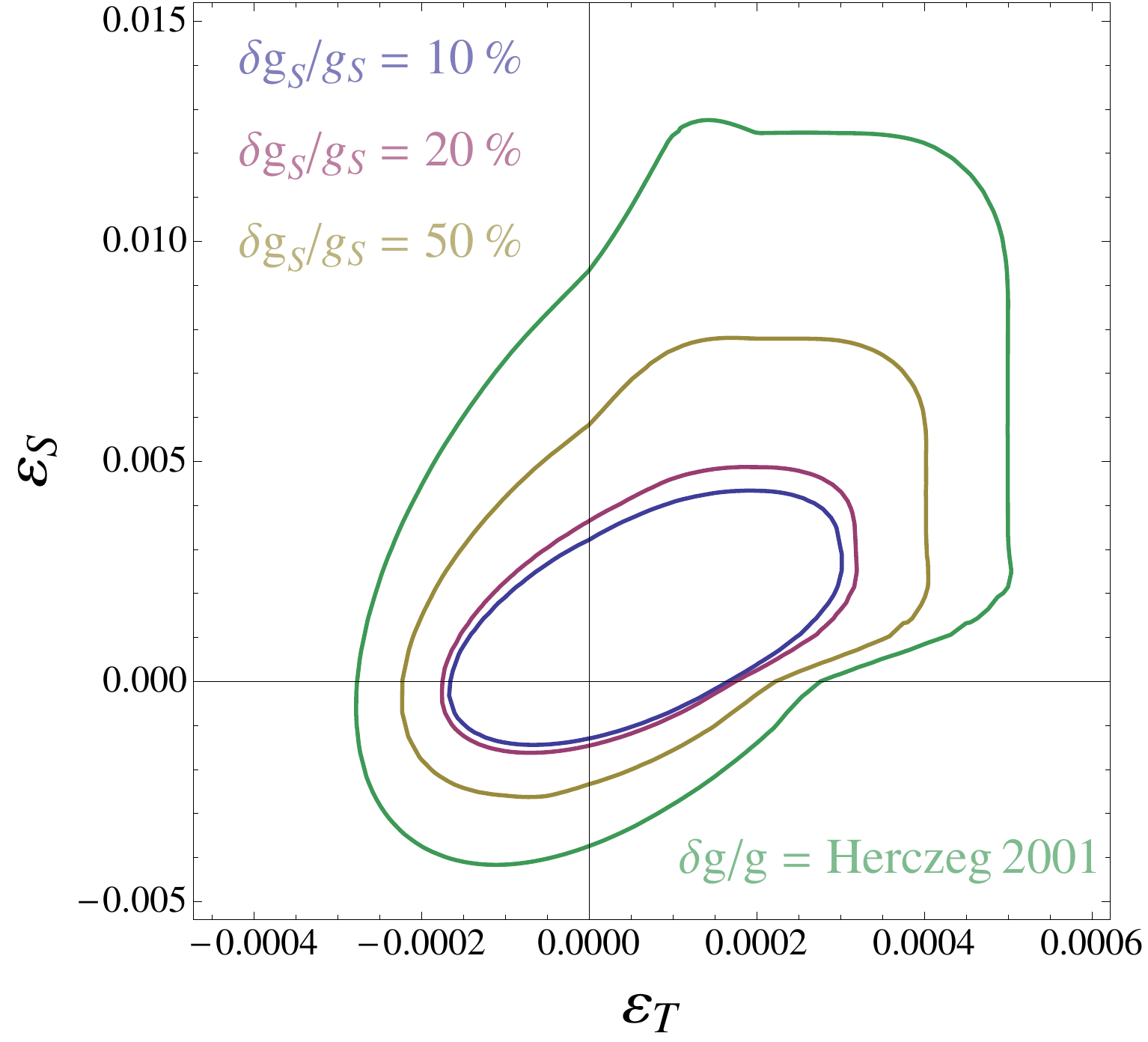}
\caption{
  \label{fig:contours}
  Constraints over the allowed regions in the $\epsilon_S, \epsilon_T$
  plane as a function of the uncertainty in $g_S$ and $g_T$ for
  different values of $\delta g_S/g_S$ assuming $\delta g_T = 2/3
  \delta g_S$. The outer contour is based on the uncertainties in
  $g_S$ and $g_T$ used in the phenomenological analysis by
  Herczeg~\cite{Herczeg2001vk}.  The convergence of the contours show that for an
  experimental uncertainty of $10^{-3}$, the goal of LQCD calculations
  should be to calculate $g_S$ and $g_{T}$ to 10-20\% accuracy.  }
\end{center}
\end{figure}

In Ref.~\cite{Bhattacharya:2011}, we show that to interpret
experimental results at the $10^{-3}$ level requires estimating $g_S$
and $g_T$ with 10--20\% precision. In Fig.~\ref{fig:contours}, taken
from Ref.~\cite{Bhattacharya:2011} and discussed further in the talk
by Bhattacharya~\cite{Bhattacharya:lat2011}, we show the allowed
region in the $\epsilon_{S,T}$ plane as a function of the uncertainty
in $g_S$ and $g_T$ assuming an experimental accuracy of
$10^{-3}$. Need for higher precision beyond 10--20\% in $g_S$ and
$g_T$ will require higher precision in experiments. Thus, the anticipated
experimental precision sets the goal of our LQCD calculations for the
2013-2014 timeframe.

As discussed further in the talk by Lin~\cite{HWLin:lat2011}, 10--20\%
precision in $g_S$ and $g_T$ requires control over the extrapolations
in quark mass to the physical pion mass and in lattice spacing to the
continuum limit, and non-perturbative determination of the
renormalization constants of the operators.  To control these
extrapolations requires high statistics data at three or more values
of the lattice spacing and light quark masses. To satisfy these
requirements within the time frame of the experiments, three years,
motivated us to use the $2+1+1$ flavor dynamical gauge configurations
generated by the MILC Collaboration using the HISQ action. At the end
of 2011, nine ensembles, each of about $1000$ lattices, at the three
lattice spacings $0.12$, $0.09$ and $0.06$ fm and each with the light 
quark mass corresponding to a pion mass of $310$, $220$ and $140$ MeV
had already been generated. Furthermore, the generation of these
lattices is continuing and supported by the USQCD national program.

For valence quarks we chose to use the improved clover fermion action
since the signal in baryon correlators with staggered fermions is
poor. Using clover valence quarks on HISQ lattices, however,
introduces the well-known problem of exceptional configurations. These
are configurations in which the clover Dirac operator has near zero
modes. When the valence and sea quark actions are the same, such
configurations are suppressed by a concommitant vanishing of their
weight in the path integral. Unchecked, exceptional configurations 
would make anomalously large contributions and degrade the reliability
of the ensemble average.  In our calculations with clover-on-HISQ,
such a cancellation is not guaranteed and checks are required for each
ensemble. Our tests, to date, on ensembles with pion masses of $310$
and $220$ MeV, do not show evidence of exceptional configurations.
Preliminary simulations at $140$ MeV suggest that this fortutious
circumstance will most likely not hold at the physical pion mass, so
we have restricted our calculations to quark masses with $310$ and
$220$ MeV pions. The analysis of the data presented by
Lin~\cite{HWLin:lat2011} shows that data at $310$ and $220$ MeV already
reduces the uncertainty due to the chiral extrapolation very
significantly.  Nevertheless, in light of the problem with 
exceptional configurations, our long-term strategy for simulations
at the physical quark mass is to perform clover-on-clover
calculations.

Another potential source of systematic errors in LQCD calculations is
guaranteeing that the matrix elements are calculated between ground
states of the proton and the neutron. Mixing with excited states
arises because lattice operators automatically couple to the
ground state and all its excited states with the same quantum
numbers. There are two ways to reduce excited state
contamination---design better interpolating operators to enhance
overlap with ground state, and increase the separation between the
source and sink. We are addressing this problem by using smeared
sources, doing the calculation with multiple source-sink separations
and by explicitly including possible contributions of excited states
in our analysis.

Our existing calculations on the $0.12$ and $0.09$ fm lattices, with
pion masses of $310$ and $220$ MeV and 400--500 configurations, show
that the statistical errors in $g_S$ are 5--6 times those in $g_T$ and
$g_A$. The uncertainty in the latter is already at the 5--10\% level.
These preliminary studies also show that there is very small
dependence of the estimates on the lattice spacing.  This gives us
confidence that a reliable continuum extrapolation will be possible
using data obtained at $0.12$, $0.09$ and $0.06$ fm.  We are therefore
confident that analyzing the full set of $1000$ lattices in each
ensemble will yield estimates with 10--20\% accuracy in $g_S$ and
$g_T$. We are also on schedule to reach this milestone over the next
two years, in time when the first experimental results are expected.

\section*{Acknowledgments}
We thank our collaborators in the theory effort (V. Cirigliano,
A. Filipuzzi, M. Gonzalez-Alonso and M. Graesser) and our experimental
colleagues. We thank the MILC Collaboration for sharing the HISQ
lattices.  These calculations were performed using the Chroma software
suite~\cite{Edwards:2004sx}.  Numerical simulations were carried out
in part on facilities of the USQCD Collaboration, which are funded by
the Office of Science of the U.S. Department of Energy, and the
Extreme Science and Engineering Discovery Environment (XSEDE), which
is supported by National Science Foundation grant number OCI-1053575.
The speaker is supported by the DOE grant DE-KA-1401020.

\bibliographystyle{apsrev}
\bibliography{2011_LatProc_Gupta}
\end{document}